\documentclass[12pt]{iopart}

\usepackage{times}
\usepackage{graphicx}
\usepackage{subfigure}
\usepackage{amssymb}
\usepackage{amsfonts}

\newcommand{\beq}{\begin{equation}}
\newcommand{\eeq}{\end{equation}}

\newcommand{\ba}{\begin{eqnarray}}
\newcommand{\ea}{\end{eqnarray}}

\def\L{\Phi}

\def\gs{\mathrel{\lower0.6ex\hbox{$\buildrel {\textstyle >}\over{\scriptstyle \sim}$}}}
\def\ls{\mathrel{\lower0.6ex\hbox{$\buildrel {\textstyle <}\over{\scriptstyle \sim}$}}}

\begin{document}

\title[Spherical symmetry and dark energy]{Spherical symmetry in a dark energy permeated space-time}

\author{
        Ninfa Radicella$^{1}$,
        Mauro Sereno$^{2,3}$,
        Angelo Tartaglia$^{2,3}$
        }

\address{$^1$ Departament de F\'{\i}sica, Universitat Aut\`{o}noma de Barcelona}
\address{$^2$ Dipartimento di Fisica, Politecnico di Torino, Corso Duca degli Abruzzi 24, 10129 Torino, Italia}
\address{$^3$ INFN, Sezione di Torino, Via Pietro Giuria 1, 10125, Torino, Italia}
\ead{ninfa.radicella@uab.cat,
mauro.sereno@polito.it,
angelo.tartaglia@polito.it}

\begin{abstract}
The properties of a spherically symmetric static space-time permeated of dark energy are worked out. Dark energy is viewed as the strain energy of an elastically deformable four dimensional manifold. The metric is worked out in the vacuum region around a central spherical mass/defect in the linear approximation. We discuss analogies and differences with the analogue in the de Sitter space time and how these competing scenarios could be differentiated on an observational ground. The comparison with the tests at the solar system scale puts upper limits to the parameters of the theory, consistent with the values obtained applying the classical cosmological tests.
\end{abstract}




\maketitle

\section{Introduction}

The general relativity (GR)\ theory has given to space-time a physical
status which makes of it one of the basic ingredients of the universe, being
the other matter/energy, however usually the nature of space-time is not
really given much attention. In the, till now, unsuccessful attempts to
quantize the gravitational field, space-time is in practice conceived as a
field, much like as for the other interactions and for matter. Despite the
enormous efforts spent on the front of quantum gravity \cite{loll98,
rovelli08, marteens10}, both in the string theory and in the loop quantum
gravity approaches, and notwithstanding the undoubtable progress and
hindsights obtained with the mathematical machinery of those theories, the
main questions still resist answers that can be both globally consistent and
unambiguously verifiable.

On the other hand, while quantum gravity tries to solve fundamental problems
at the smallest scales and the highest energies, a problem also exists at
large scales where classical approaches are in order. Observation \cite{riess98, komatsu10} has forced people to hypothetically introduce in the
universe entities that have scarse or no reference to the matter/energy we
know by experiment at intermediate or small scales. We apparently need dark
matter and dark energy \cite{kamionkowski}, and, especially for the second,
when trying to work out its properties and to build some physical
interpretation of its nature, people are led to results which, to say the
least, are far away from our intuition and experience .

Another approach consists in trying to modify the general theory of
relativity \cite{peebles03,dvali00,sotiriou08}, outside
and beyond the simplicity criteria that, despite the mathematical
complexity, guided its development. Both the dark-something and the modified
GR theories are in a sense ad hoc presciptions. Preserving an internal
consistency requirement the theories look for Lagrangians for the universe
apt to yield equations reproducing or mimicking what we observe.

The approach we have already followed in previous works \cite{CQG} consists
in treating space-time as a classical four-dimensional continuum behaving as
three-dimensional material continua do \cite{landau, eshelby}. An
appropriate name for the theory worked out in this way is Strained State
Theory (SST) since the new features it introduces are contained in the
strain tensor expressing the difference between a flat undifferentiated
four-dimensional Euclidean manifold and the actual space-time with its
curvature, originated from matter/energy distributions as well as from
texture defects in the manifold as such. In a sense SST is a theory of the
dark energy where the latter is a vacuum deformation energy present when the
space-time manifold is curved.

Here we shall discuss the behavior of such a strained space-time when some
external cause (be it a mass or a defect) induces a spherical symmetry in
space. In a sense we will treat the analog of the Schwarzschild problem in a
dark energy permeated environment.

As it will result, the presence of the strain energy appears at the cosmic scale, without affecting in a sensible way the physics at the scale of the solar system. In any case the data from the solar system will constrain the value of the parameters of the theory. Since the solution of the problem will be
attained by an approximation method, the asymptotic region, where the effect
of strain would be dominant, will be excluded from our description.

\section{The strained state of space-time}

The essence of the strained state theory is in the idea that space-time is a
four-dimensional manifold endowed with physical properties similar to the
ones we know for deformable three-dimensional material continua. In practice
we may think that our space-time, which we shall call the natural manifold,
is obtained from a flat four-dimensional Euclidean manifold, which will be
our reference manifold. The deformation, i.e. the curvature, of space-time
is due to the presence of matter fields as in GR or to the presence of
texture defects in the manifold, however here we assume that space-time
resists to deformation more or less as ordinary material continua do. In
practice, according to this approach, we introduce in the Lagrangian density
of space-time, besides the traditional Einstein-Hilbert term, an "elastic
potential term" built on the strain tensor in the same way as for the
classical elasticity theory. The additional term in a sense accounts for the
presence of a dark energy or even "curvature fluid" \cite{capo}.
The bases of SST are described in ref. \cite{CQG}; here we review the
essential.

The complete action integral of the theory is%
\begin{equation}
S=\int \left( R+\frac{1}{2}\left( \lambda \varepsilon ^{2}+2\mu \varepsilon
_{\mu \nu }\varepsilon ^{\mu \nu }\right) +\mathcal{L}_{matter}\right) \sqrt{%
-g}d^{4}x  \label{action}
\end{equation}

Of course $R$ is the scalar curvature of the manifold; the parameters $\lambda $ and $\mu $ are the Lam\'{e} coefficients of
space-time; $\varepsilon _{\mu \nu }$ is the strain tensor of the natural
manifold and $\varepsilon =\varepsilon _{\alpha }^{\alpha }$; $\mathcal{L}_{matter}$ is the Lagrangian density of matter/energy. The strain
tensor is obtained by comparison of two corresponding line elements, one in
the natural frame and the other in the reference frame. By definition it is%
\begin{equation}
\varepsilon _{\mu \nu }=\frac{1}{2}\left( g_{\mu \nu }-E_{\mu \nu }\right)
\label{strain}
\end{equation}%
where $g_{\mu \nu }$ is the metric tensor of the natural manifold and $%
E_{\mu \nu }$ is the Euclidean metric tensor of the reference frame.

The action (\ref{action}) has already been used both in ref. \cite{CQG} and
\cite{cosmo} in order to describe the accelerated expansion of the universe,
and has given positive results when tested against four typical cosmological
tests \cite{cosmo}.

\section{Spherical symmetry in space.}

Now we focus on a stationary physical system endowed with spherical symmetry
in space. Of course there must be a physical reason for the symmetry to be
there, which means that "something" must exist in the central region of the
space-time we are considering. This can be either a time independent
spherical aggregate of mass/energy or a line defect\footnote{%
Line defect refers to the full four-dimensional space-time and the line will
be time-like, so that in space the defect will appear to be pointlike.}. The
general form of the line element of a space-time with the given symmetry is
well known:%
\begin{equation}
ds^{2}=fd\tau ^{2}-hdr^{2}-r^{2}\left( d\theta ^{2}+\sin ^{2}\theta d\phi
^{2}\right)  \label{linen}
\end{equation}%
where $f$ and $h$ are functions of $r$ only and Schwarzschild coordinates
have been used.

The corresponding line element in the flat Euclidean reference frame will be:
\begin{equation}
ds_{r}^{2}=d\tau ^{2}+\left( \frac{dw}{dr}\right) ^{2}dr^{2}+w^{2}\left(
d\theta ^{2}+\sin ^{2}\theta d\phi ^{2}\right)  \label{linef}
\end{equation}

\begin{figure*}
\centering
	\includegraphics[width=0.6\textwidth]{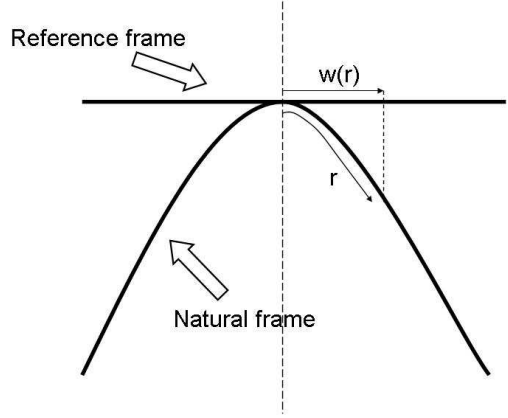}
	\caption{When using the coordinates of the natural frame $r$, the radial coordinate of the reference frame is a function $w\left( r\right) $ depending on the actual curvature of the natural frame.}
	\label{fig1}
\end{figure*}

In principle we have four degrees of freedom (together with the flatness
condition) in the choice of the coordinates on the reference manifold,
however when we decide to evidence the same symmetry as the one present in
the natural frame, the gauge functions in practice reduce to one. This is
the meaning of the $w$ function, only depending on $r$, in eq. (\ref{linef}). Fig.~\ref{fig1} pictorially clarifies the role of the gauge function.

By direct inspection of formulae (\ref{linen}) and (\ref{linef}) and using the definition (\ref{strain}) we can easily read out the non-zero elements of the strain tensor for this physical configuration:
\begin{eqnarray}
\varepsilon _{00} & =& \frac{f-1}{2} \\
\varepsilon _{rr} & =& -\frac{h+w^{\prime 2}}{2} \\
\varepsilon _{\theta \theta }& =& -\frac{r^{2}+w^{2}}{2} \\
\varepsilon _{\phi \phi }& =& -\frac{r^{2}+w^{2}}{2}\sin ^{2}\theta%
\end{eqnarray}

From now on, primes will denote derivatives with respect to $r$.

Once we have the strain tensor, we are able to write the contribution to the
Lagrangian density of space-time due to the strain present in the natural
manifold. The needed ingredients are:%
\begin{equation}
\varepsilon =g^{\alpha \beta }\varepsilon _{\alpha \beta }=\frac{f-1}{2f}+%
\frac{h+w^{\prime 2}}{2h}+\frac{r^{2}+w^{2}}{r^{2}}  \label{trace}
\end{equation}%
and%
\begin{equation}
\varepsilon _{\alpha \beta }\varepsilon ^{\alpha \beta }=g^{\alpha \mu
}g^{\beta \nu }\varepsilon _{\alpha \beta }\varepsilon _{\mu \nu }=\frac{%
\left( f-1\right) ^{2}}{4f^{2}}+\frac{\left( h+w^{\prime 2}\right) ^{2}}{%
4h^{2}}+\frac{\left( r^{2}+w^{2}\right) ^{2}}{2r^{4}}  \label{second}
\end{equation}

For completeness let us remind that it is%
\begin{equation}
R=-\left( \frac{2}{r^{2}}-\frac{2}{hr^{2}}-\frac{f^{\prime \prime }}{fh}+%
\frac{f^{\prime 2}}{2f^{2}h}+\frac{1}{2fh^{2}}f^{\prime }h^{\prime }-\frac{2%
}{fhr}\allowbreak f^{\prime }+\frac{2h^{\prime }}{h^{2}r}\right)
\label{curvature}
\end{equation}

and%
\begin{equation}
\sqrt{-g}=\sqrt{fh}r^{2}\sin \theta  \label{det}
\end{equation}

Going back to eq. (\ref{action}) we are now able to write the full explicit
Lagrangian density of our strained space-time, with the built in
Schwarzschild symmetry. We are interested in empty space-time so in the
region we shall be considering it will be $\mathcal{L}_{matter}=0$.

From the Lagrangian density, applying the usual variational procedure, we can obtain the Euler-Lagrange equations for the $f$, $h$ and $w$ functions. The effective Lagrangian density (modulo a $\sin \theta $) is:
\begin{eqnarray}
\mathfrak{L}& =& -\left( \frac{2}{r^{2}}-\frac{2}{hr^{2}}+\frac{2h^{\prime }}{h^{2}r}\right) \sqrt{fh}r^{2}  \nonumber \\
& +& \frac{\lambda }{2}\left( \frac{f-1}{2f}+\frac{h+w^{\prime 2}}{2h}+\frac{%
r^{2}+w^{2}}{r^{2}}\right) ^{2}\sqrt{fh}r^{2}  \label{lagrangiana} \\
& +& \mu \left( \frac{\left( f-1\right) ^{2}}{4f^{2}}+\frac{\left( h+w^{\prime
2}\right) ^{2}}{4h^{2}}+\frac{\left( r^{2}+w^{2}\right) ^{2}}{2r^{4}}\right)
\sqrt{fh}r^{2}  \nonumber
\end{eqnarray}

The second derivative appearing in Eq.~(\ref{curvature}) has been eliminated by means of an integration by parts.

The $w$ function is treated as $f$ and $h$, which means that we assume it has to satisfy Hamilton's principle just as the others do. The reason for this choice is in that we are representing the correspondence between the natural and the reference manifolds as being established by an actual physical deformation process, which is something else from the obvious freedom in the choice of the coordinates. The three explicit final equations are:
\begin{eqnarray}
0& =& h -1+r\frac{h^{\prime }}{h}+\allowbreak \frac{1}{16f^{2}h}\lambda
r^{2}\left( 2fh\frac{w^{2}}{r^{2}}+4fh+3h+fw^{\prime 2}\right) \nonumber \\
&\times& \left( h-4fh-2fh\frac{w^{2}}{r^{2}}-fw^{\prime 2}\right) \label{uno} \\
& -& \frac{1}{8hf^{2}}\mu r^{2}\left( 2fh^{2}+4f^{2}h^{2}+2f^{2}h^{2}\frac{w^{4}%
}{r^{4}}-3h^{2}+f^{2}w^{\prime 4}+4f^{2}h^{2}\frac{w^{2}}{r^{2}}%
+2f^{2}hw^{\prime 2}\right)  \nonumber \\
0&=&h-1-\frac{1}{f}rf^{\prime } -\frac{1}{16hf^{2}}\lambda r^{2}\left( h-4fh-2fh\frac{w^{2}}{r^{2}}+3fw^{\prime 2}\right) \nonumber \\
&\times & \left( h-4fh-2fh\frac{w^{2}}{r^{2}}-fw^{\prime 2}\right)  \label{due} \\
&-&\frac{1}{8hf^{2}}\mu r^{2}\left( h^{2}+4f^{2}h^{2}+2f^{2}h^{2}\frac{w^{4}}{r^{4}}-2fh^{2}-3f^{2}w^{\prime 4}+4f^{2}h^{2}\frac{w^{2}}{r^{2}}-2f^{2}hw^{\prime 2}\right) \nonumber \\
0&=&\frac{\lambda }{2fh^{2}}w^{\prime \prime }\left( hr^{2}-3fr^{2}w^{\prime2}-4fhr^{2}-2fhw^{2}\right)  \nonumber\\
& -& \frac{\lambda }{h}ww^{\prime 2}-\frac{\lambda r}{h^{2}}\left( \frac{f^{\prime }}{4f}r-\frac{3h^{\prime }}{4h}r+1\right) w^{\prime 3} \nonumber \\
& +& \lambda \frac{w^{\prime }}{h}\left( \left( -\frac{1}{2}w^{2}-r^{2}-\frac{1}{4f}\allowbreak r^{2}\right) \frac{f^{\prime }}{f}+\left( r^{2}+\frac{1}{2}w^{2}-\frac{1}{4f}r^{2}\right) \frac{h^{\prime }}{h}+\frac{1}{f}r-4r\right) \nonumber \\
&+&\lambda w\left( 4+\frac{2}{r^{2}}w^{2}-\frac{1}{f}\right) +\mu \frac{r^{2}}{h^{2}}w^{\prime \prime }\left( -3w^{\prime 2}-h\right) \nonumber \\
&-&\frac{\mu }{h^{2}}\left( 2r-\frac{3}{2h}r^{2}h^{\prime }+\frac{1}{2f}r^{2}f^{\prime }\right) w^{\prime 3} \nonumber \\
&+& \mu \frac{r}{h}\left( \frac{h^{\prime }}{2h}r-\allowbreak 2-\frac{f^{\prime }}{2f}r\right) w^{\prime }+2w\mu \left( 1+\frac{w^{2}}{r^{2}}\right) \label{tre}
\end{eqnarray}

As it is immediately seen, the three equations are highly non-linear, first
order differential in $f$ and $h$, second order differential in $w$. Solving
them exactly is apparently a desperate task, but we shall see that it is
possible to proceed perturbatively.

\section{Approximate solutions}

Looking at eqs. (\ref{uno}) and (\ref{due}) we see that there are a number
of terms multiplying either the $\lambda $ or $\mu $ parameter, while others
do not. From the application of the theory to the cosmic expansion we know
that the values of $\lambda $ and $\mu $ are indeed very small \cite{CQG}%
\cite{cosmo}; the dimension of the parameters is the inverse of the square
of a length, so we may say that for distances small with respect to some
typical radius $\tilde{r}$ the products $\lambda r^{2}$ and $\mu r^{2}$ will
be much smaller than $1$. The typical $\tilde{r}$ is $\sim 10^{26}$ m $\sim
10^{4}$ Mpc \cite{CQG}\cite{cosmo}.

We are then led to solve the equations by successive approximations. Our
first step in the approximation process will be to neglect the terms
multiplying $\lambda $ and $\mu $\footnote{%
For simplicity we assume that $\lambda $ and $\mu $ are of the same order of
magnitude.} so that the zero order equations become:
\begin{equation}
\allowbreak h_{0}-1+r\frac{h_{0}^{\prime }}{h_{0}}=0  \label{uno0}
\end{equation}%
$\allowbreak $%
\begin{equation}
h_{0}-1-r\frac{f_{0}^{\prime }}{f_{0}}=0.  \label{due00}
\end{equation}

The solution is the typical Schwarzschild one:%
\begin{equation}
f_{0}=1-2\frac{m}{r}  \label{Schwf}
\end{equation}%
\begin{equation}
h_{0}=\frac{1}{f_{0}}=\frac{1}{1-2\frac{m}{r}}  \label{Schwh}
\end{equation}
Looking to the recovery of the Newtonian limit we see of course that the
integration constant does actually coincide with the central mass $m$.

In practice we can write that the solutions of equations Eqs.~(\ref{uno},\ref{due},\ref{tre}) are of the type

\begin{eqnarray}
f &=&f_{0}+\phi  \nonumber \\
h &=&h_{0}+\chi  \label{linsol} \\
w &=&lr\left( 1+\psi \right)  \nonumber
\end{eqnarray}%
with $\phi $, $\chi $, $\psi <<1$. Up to this moment we have not said
anything about the relative size of $m/r$ with respect to the $\lambda r^{2}$
or $\mu r^{2}$ terms, inside the fiducial radius $\tilde{r}$. We know
however that, outside any Schwarzschild horizon, it is $m/r<1$ so that any $%
m\lambda r$ or $m\mu r$ term will be smaller that the $\lambda r^{2}$ or $%
\mu r^{2}$ terms. On these bases we conclude that at the lowest
approximation order $\phi $, $\chi $ and $\psi $ are functions of $\lambda
r^{2}$ and $\mu r^{2}$.

The adimensional scale factor $l$ would be arbitrary in a trivial flat
space-time, but this is not the case here.

Introducing the developments (\ref{linsol}) into (\ref{uno}) and (\ref{due})
and keeping the terms up to the first order in $\lambda r^{2}$ and $\mu
r^{2} $ we see that only $w=lr$ plays a role, so that we do not need to
worry about the unknown function $\psi$. In any case the functional form of $w$ is determined by requiring that in absence of elastic deformation the reference metric be Euclidean, which suggests that $\psi$ in Eq.~(\ref{linsol}) must go to zero for $\lambda=\mu=0$. We nevertheless explored the possibility that a different ansatz for $w$ could bring a new set of solutions; we considered as functional forms for $w$ either Maclaurin or Taylor expansions in (inverse) powers of $r$ and we found that higher order terms in the expansion must zero out. The linear $r$ term considered in Eq.~(\ref{linsol}) is then the only relevant one.

Finally we obtain:
\begin{eqnarray}
\phi &=&\Phi r^{2}  \label{soluzphi} \\
\chi &=&\Psi r^{2}  \label{soluzchi}
\end{eqnarray}

The explicit expressions of the $\Phi $ and $\Psi $ parameters are:
\begin{eqnarray}
\Phi & =& \frac{\lambda }{16}\left( 3l^{4}+2l^{2}-1\right) +\frac{\mu }{8} \left( l^{4}-1\right) \\
\Psi & =& \frac{\lambda }{16}\left( 3l^{4}+10l^{2}+7\right) +\frac{\mu }{8} \left( l^{2}+1\right) ^{2} \label{lambdapsi}
\end{eqnarray}

The result does indeed depend on the value of $l$; different values
correspond to different situations. We shall comment on this in a while. In
any case it is $\Phi \neq \Psi $ unless $\mu =-2\lambda $.

We could also have started from pure flat space-time as zero order
approximation, but at the end we would have found again the same solution,
i.e. Schwarzschild plus (\ref{soluzphi}) and (\ref{soluzchi}).

\section{The metric tensor}

Explicitly writing the results found in the previous section we see that we have different regions with specific approximate forms for the line element. Cosmological constraints suggest that $\lambda \sim \mu \sim 10^{-52} \mathrm{m}^{-2}$. Then, for masses as large as those of galaxies or clusters of galaxies, we can distinguish three regimes. An internal region, where $1>>m/r>>\lambda r^{2}$, $\mu r^{2}$:

\begin{equation}
ds^{2}\simeq \left( 1-2\frac{m}{r}+\Phi r^{2}\right) d\tau ^{2}-\left(
\frac{1}{1-2\frac{m}{r}}+\Psi r^{2}\right) dr^{2}-r^{2}\left( d\theta
^{2}+\sin ^{2}\theta d\phi ^{2}\right)  \label{lineaz}
\end{equation}

An intermediate region, where $1>>m/r\sim \lambda r^{2},\mu r^{2}$:%
\begin{equation}
ds^{2}\simeq \left( 1-2\frac{m}{r}+\Phi r^{2}\right) d\tau ^{2}-\left( 1+2%
\frac{m}{r}+\Psi r^{2}\right) dr^{2}-r^{2}\left( d\theta ^{2}+\sin
^{2}\theta d\phi ^{2}\right)  \label{linea}
\end{equation}

An outer region\allowbreak , where $r<\tilde{r}$ but $1>>\lambda r^{2},\mu
r^{2}>>m/r$:

\begin{equation}
ds^{2}\simeq \left( 1+\Phi r^{2}\right) d\tau ^{2}-\left( 1+\Psi
r^{2}\right) dr^{2}-r^{2}\left( d\theta ^{2}+\sin ^{2}\theta d\phi
^{2}\right)  \label{lineae}
\end{equation}

Our approximate solutions are unfit to describe the asymptotic region where $\lambda r^{2}$,$\mu r^{2}\sim 1$ or bigger. This is the cosmological domain and the problem opens of the embedding in a given cosmic background
space-time.

The internal metric has vanishing values of $g_{00}$ for
\begin{equation}
r_{00} \simeq \sqrt[3]{\sqrt{\frac{m^{2}}{\Phi ^{2}}+\frac{1}{27\Phi ^{3}}}+%
\frac{m}{\Phi }}-\frac{\frac{1}{3\Phi }}{\sqrt[3]{\sqrt{\frac{m^{2}}{%
\Phi ^{2}}+\frac{1}{27\Phi ^{3}}}+\frac{m}{\Phi }}}
\end{equation}%
whose limit correctly goes to $2m$ when $\Phi \rightarrow 0$.

Eq. (\ref{lineae}) holds also in the case of a defect without mass. In that
case the scalar curvature in the inner region, to first order in $\lambda
r^{2}$,$\mu r^{2}$, is:

\begin{equation}
R\simeq 6\allowbreak \left( \Psi -\Phi \right)  \label{curva}
\end{equation}%
$\allowbreak $

Explicitly it is:

\begin{equation}
R\simeq 3\left( \allowbreak 1+l^{2}\right) \left( \lambda +\frac{1}{2}\mu
\right)
\end{equation}%
The curvature is a scalar quantity, independent from the coordinates. As we
see the result depends on $l$ so that we are forced to attach a physical
meaning to that parameter. Since we are now treating a mass-free situation
we are led to conclude that some defect is present in the origin and its
relevance is quantitatively expressed by the value of $l$. Another remark is
that the curvature in the origin, even in the absence of mass, is never
zero, if we only allow for real values of $l$: the initial Euclidean
reference frame can be brought to locally coincide with a Minkowskian
tangent space only for imaginary values of $l$, in which case actually the
initial frame would have been Minkowskian.

\section{Perihelion precession}
\label{sec:peri}

\begin{table}
\centering
\caption{\label{tab_plan} Limits on $\Phi$ due to extra-precession of the inner planets of the solar system. Extra-precession values $\delta\dot{\omega}$ are from \cite{fie11}.}
\begin{tabular}{l|c|c}
Name  &  $\delta\dot{\omega}$ [mas/year] & $\Phi~[\mathrm{m}^{-2}] $ \\
\hline
\hline
Mercury  &  $0.6$   &  $\ls 0.6 {\times} 10^{-40}$
\\
Venus    &  $1.5$     &  $\ls 0.6 {\times} 10^{-40}$
\\
Earth    &  $0.9$  &    $\ls 0.2{\times} 10^{-40}$
\\
Mars &      $0.15$  & $\ls 0.2{\times} 10^{-41}$
\\
Jupiter &      $42$  & $\ls 0.8{\times} 10^{-40}$
\\
Saturn &      $0.65$  & $\ls 0.5{\times} 10^{-42}$
\\
\end{tabular}
\end{table}

Precessions of the perihelia of the Solar system planets have provided stringent local tests for competing theories of gravity \cite{isl83,je+se06,se+je06b}. A metric deviation of the form $\delta g_{00} \simeq \Phi r^2$ from the standard result obtained in general relativity induces a precession angle after one orbital period of
\begin{equation}
\Delta \phi \simeq 3 \pi \Phi \frac{s^3}{ r_\mathrm{g}} (1-e^2)^{1/2} .
\end{equation}
where $\Delta \phi $ is in radians; $s$ and $e$ are the semi-major axis and the eccentricity of the unperturbed orbit, respectively, and, $r_\mathrm{g} = G M/c^2$ is the gravitational radius of the central body.

Data from space flights and modern astrometric methods make it possible to create very accurate planetary ephemerides and to precisely determine orbital elements of Solar system planets \cite{pit05,fie11}. Results are compatible with GR predictions, so that any effect induced by modifications of the gravity law may be to the larger extent of the order of the statistical uncertainty in the measurement of the precession angle. Here we consider the planetary ephemerides in \cite{fie11}.

The accurate measurement of Saturn perihelion shift provides the tighter bound on $\L$ from solar system tests, $\Phi \ls 0.5\times10^{-42}  \mathrm{m}^{-2}$, see Table~\ref{tab_plan}. Local tests on perihelion precession put bounds on $\Phi$, whereas cosmological observations constrain a different combination of parameters of the CD theory, the $B [\equiv (\mu/4) (2 \lambda+\mu)/(\lambda + 2 \mu) ] $ parameter in \cite{cosmo}. Local bounds are anyway nine orders of magnitude less constraining than cosmological tests. Other solar or stellar system tests can probe gravitational theories but they are usually less constraining than results from measurements of the precession angle of the planets in the inner Solar system \cite{se+je06a}.

\section{Radial acceleration}

Another interesting quantity is the radial acceleration of an observer
instantaneously at rest. Now we refer to the geodetic equations deducible
from line element (\ref{lineaz}). Being interested to a pure radial fall, we
put $d\theta /ds=d\phi /ds=0$; the remaining pair of equations is:

\begin{equation}
\begin{array}{c}
\frac{d^{2}\tau }{ds^{2}}+2\left( r\Phi +\frac{m}{r\left( r-2m\right) }%
\right) \frac{d\tau }{ds}\frac{dr}{ds}\simeq 0 \\
\frac{d^{2}r}{ds^{2}}+\left( r\Phi +\frac{m}{r^{3}}\left( r-2m\right)
\right) \left( \frac{d\tau }{ds}\right) ^{2}+\left( r\Psi -\frac{m}{r\left(
r-2m\right) }\right) \left( \frac{dr}{ds}\right) ^{2}\simeq 0%
\end{array}
\label{geod}
\end{equation}

For a momentarily fixed position it is also $dr/ds=0$, so that the equations
become:

\begin{equation}
\begin{array}{c}
\frac{d^{2}\tau }{ds^{2}}\simeq 0 \\
\frac{d^{2}r}{ds^{2}}+\left( r\Phi +\frac{m}{r^{3}}\left( r-2m\right)
\right) \left( \frac{d\tau }{ds}\right) ^{2}\simeq 0%
\end{array}
\label{geod1}
\end{equation}

Let us evaluate the proper radial acceleration; we see that

\begin{equation}
\frac{d^{2}r}{d\tau ^{2}}\simeq -\frac{m}{r^{2}}\left( 1-2\frac{m}{r}\right)
-r\Phi   \label{accelerre}
\end{equation}

The strained state of space-time adds a contribution to the Newtonian and post Newtonian acceleration strengthening (weakening) the force of gravity for a positive (negative) value of $\Phi$.

An additional term in the form of Eq.~(\ref{accelerre}) causes a change in Kepler's third law. Because of $\Phi$, the radial motion of a test body around a central mass $M$ is affected by an additional acceleration which perturbs the mean motion. For a radial acceleration in the form of $\Phi r$ perturbing an otherwise Newtonian orbit, the mean motion $n=\sqrt{G M/s^3}$ is changed by \cite{se+je06a}
\begin{equation}
\frac{\delta n}{n}= -\Phi\frac{s^3}{r_\mathrm{g}}.
\end{equation}

In principle, the variation of the effective gravitational force felt by the solar-system inner planets with respect to the effective forces felt by outer planets could probe new physics. However, observational uncertainties on the mean motion, i.e. on the measured semi-major axis of the solar-system planets, are quite large \cite{pit05}. The tighter constraint comes from the Earth orbit, whose orbital axis is determined with an accuracy of $\delta s = 0.15~\mathrm{m}$ \cite{pit05}. This provides an upper bound to $\Phi$ of the order of $\ls 0.2 \times 10^{-40}~\mathrm{m}^{-2}$.

\section{Matching with the Robertson-Walker metric}

Up to now, we only required the metric to be spherically symmetric. The homogeneous and isotropic space-time is then a particular case of our local analysis. This highly symmetric case is obtained by considering a manifold without a central mass, i.e., $m=0$, and with just a central defect that can force the space-time to be homogeneous too. This condition can fix the size $l$ of the defect. It can be then interesting to compare with the exact solutions obtained with Robertson and Walker coordinates in the cosmological case. Being our new result local, we have to consider the RW metric at the present time. The today value of the curvature is
\begin{equation}
R_\mathrm{RW} = 12 B \left(1-\frac{1}{a_\mathrm{0}^2}\right),
\end{equation}
where $a_\mathrm{0}$ is the present value of the scale factor and $B \equiv (\mu/4) (2 \lambda+\mu)/(\lambda + 2 \mu)$. We can then look for the size $l$ of the central defect such that the resulting space-time is isotropic and homogeneous at the same time by requiring that the local value of the curvature is equal to the value in the RW metric. We get
\begin{equation}
l \simeq \frac{1}{a_\mathrm{0}} \sqrt{\frac{2\mu -a_\mathrm{0}^2 (\lambda+4\mu)}{\lambda +2\mu}}.
\end{equation}
In \cite{CQG}, cosmological expansion was explained as a consequence of a defect in an elastic medium. The above result describes the today expansion factor in terms of the local size of the defect.

\section{Comparison with massive gravity}

The SST theory looks very similar to the classical massive gravity theory initially proposed by Fierz and Pauli (FP) \cite{fierz}. At first sight indeed our Lagrangian corresponds to the FP one; if the similarity were an actual coincidence we would have to face the same kind of inconveniences which are known to plague massive gravity. These are essentially the so called van Dam-Veltman-Zakharov (vDVZ) discontinuity \cite{van}\cite{zakh} and the presence of ghosts appearing to various orders. In another work \cite{libro} one of us already had considered the problem and the remark had been that the FP theory is based on a first order perturbative treatment on a flat Minkowskian background; this is not the case of the SST which is "exact" and does not assume that the elements of the strain tensor are small. However the interest in massive gravity has stimulated a vast effort to formulate a theory valid to all orders and free from the mentioned troubles; a good review of the progress along the mentioned search can be read in ref. \cite{hin11} and we will refer to it for further considerations. Again when considering the non-linear version of massive gravity we find a Lagrangian which apparently corresponds to the one of SST; however, as we shall see in a moment, the two Lagrangians are different.
In fact non-linear massive gravity can be seen as a four dimensional bi-metric theory \cite{hin11}. One metric is dynamical, whereas the second is not coupled to the actual universe and is formally frozen, i.e. it describes a non-dynamical Einstein space background \cite{dam+al03}. The non-dynamical metric is used to raise and lower the indices of the $h_{\mu\nu}$ tensor which is the equivalent of our strain tensor \cite{hin11} or is combined with the full $g_{\mu\nu}$ to produce the scalars needed for the potential in \cite{dam+al03}.

In the SST theory, there is just one metric, $g_{\mu\nu}$, which is used for all tasks pertaining to a metric tensor. Our $E_{\mu\nu}$ tensor appearing in Eq.~\ref{strain} is indeed described as the metric tensor of the flat reference frame but is \textit{not} any metric at all for the natural frame. The only existing frame is the natural one; the reference frame belongs to a logically preceding phase in a descriptive paradigm where the present space-time is obtained as a deformation of some previous undeformed flat state, but the previous stage does not exist or coexist with the natural frame.  $E_{\mu\nu}$ is not used to raise or lower any index; rather the full metric $g_{\mu\nu}$ is used to raise and lower all indices including those of $E_{\mu\nu}$, which is a symmetric tensor in the natural manifold. Often we find in the literature also the claim that in massive gravity theories General Coordinate Transformation (GCT) invariance is broken by the "massive" term (see for instance ref.~\cite{GCT}) and various devices are needed in order to restore it; this is not the case of SST, since in our theory all objects are true tensors.
 The $E_{\mu\nu}$ tensor does not even coincide with the metric of the local tangent space, which is Minkowski and position depending. As a matter of fact, results in the SST theory can equally well be obtained starting from an Euclidean or a Minkowskian reference, which again indicates that the natural metric is the only relevant one.

 The difference we have pointed out tells us that there is no obvious affection of the SST by the same difficulties affecting the classical massive gravity theories. By the way the vDVZ discontinuity is indeed absent in the cosmological application of SST as well as in the case studied here, where the solutions go smoothly to GR when one lets $\lambda$ and $\mu$ go to zero. One further comment about ghosts is in order. The whole discussion of ghosts implies a field theoretical approach to gravity and/or the study of propagating perturbations. As for the former we know that gravity cannot be described as a spin-2 field on a flat background; furthermore one cannot even say that the graviton exists, so we continue to use the expression "mass of the graviton" as a sort of abbreviation for something else. Once one analyzes the perturbations the problem of negative kinetic energy is discussed order by order, but the conclusions that one can draw summing to all orders is not well defined. Various tricks have been devised in order to get rid of ghosts up to a predefined order (e.g. the fourth or the fifth \cite{gaba}). Here we do not enter into the discussion, simply stress that: a) as seen above, we have just one metric, which is a properly defined metric; b) that SST is not based on a peculiar perturbative development. When taken globally, the problems of SST, if any, are shared with the cosmological constant model of space-time.
 
Actions in either of the two theories  could be formally identified if we lower and raise indices with the full metric rather than the frozen metric in non-linear massive gravity. Then, in case the full metric and the full determinant can be expanded in powers of the deviation, we can re-organize the terms in the potential and show that the two approaches would carry the same information  \cite{hin11}. However, this analogy has been probed only with this perturbative approach and we have a direct correspondence to first order only. The SST theory is intrinsically non linear. Just as an example, the expansion technique cannot be applied in the cosmological case, that was exactly analyzed in \cite{CQG}. We can then not conclude that the SST theory suffers the same pathologies as the standard massive gravity.

The comparison of what is known in the spherically symmetric case further shows how known problems affecting massive gravity do not automatically apply to SST. Usual problems in the standard massive gravity have been discussed expanding the equations around the flat solution in terms of small functions. An alternative expansion in the squared mass, which would mimic the expansion technique used in this paper for the SST theory, might hopefully show a smooth limit without discontinuity. Some recent analytic solutions in non-linear massive gravity \cite{koy+al11} have shown a branch of exact solutions which corresponds to Schwarzschild-de Sitter space-times where the curvature scale of de Sitter space is proportional to the squared mass of the graviton. This is similar to the results found in the present paper for the SST theory. Even if these arguments are not conclusive they are nevertheless encouraging.

\section{Conclusions}

We have found the approximate configuration of the space-time surrounding a
spherical mass distribution or texture defect independent from time,
assuming that a dark energy given by the strain of the manifold is present.
As expected, we see that the strain of space-time contributes "locally"
extremely tiny corrections to the Schwarzschild solution. These corrections
lead to a slight displacement of the horizon in the inner region and to
changes of the precession rates of the periapsis of orbiting celestial
bodies as well as of the proper radial acceleration. The comparison of the
expected corrections with the data known in the solar system puts upper
bounds to the parameters of the theory which are fully consistent with the
results found applying the SST to the universe as a whole. Summing up: the
Strained State Theory, while giving a physical interpretation to the dark
energy in vacuo, accounts for the accelerated expansion of the universe and
passes other relevant cosmological tests \cite{cosmo}; locally it leads to
effects that become visible at the scale of galaxy clusters or bigger.

Our results also show differences between the local predictions of the SST theory versus the standard interpretation of dark energy as a cosmological constant. In particular, we found that in the SST $g_{00} \neq -g_{rr}^{-1}$, which is a main difference with the de Sitter metric and implies that the two competing theories are not degenerate and might be distinguished with very accurate data.

The additional term $\Phi$ to the metric element $g_{00}$ influences the gravitational potential whereas $\Psi$ contributes to the space curvature perturbation. $\Phi$ directly affects the Poisson equation and determines the modified growth of structure with respect to GR. $\Psi$ together with $\Phi$ influences the null geodesics of light and might be constrained with gravitational lensing measurements.

\ack
Ninfa Radicella was funded by the Spanish Ministry of Science and Innovation through the "Subprograma Estancia J\'ovenes Doctores Extranjeros, Modalidad B", Ref: SB2009-0056.

\section*{References}


\end{document}